
\voffset 0.1in
\hoffset 0.1in

\def\gboxit#1{\hbox{\vrule\vbox{\hrule\kern3pt\vtop
{\hbox{\kern3pt#1\kern3pt}
\kern3pt\hrule}}\vrule}}

\def\ttilde#1{\raise2ex\hbox{${\scriptscriptstyle(}\!
\sim\scriptscriptstyle{)}$}\mkern-16.5mu #1}
\def\dddots#1{\raise1ex\hbox{$^{\ldots}$}\mkern-16.5mu #1}
\def\siton#1#2{\raise1.5ex\hbox{$\scriptscriptstyle{#2}$}\mkern-16.5mu
#1}
\def\upleftarrow#1{\raise1.5ex\hbox{$\leftarrow$}\mkern-16.5mu #1}
\def\uprightarrow#1{\raise1.5ex\hbox{$\rightarrow$}\mkern-16.5mu #1}
\def\upleftrightarrow#1{\raise1.5ex\hbox{$\leftrightarrow$}\mkern-16.5mu
#1}
\def\bx#1#2{\vcenter{\hrule \hbox{\vrule height #2in \kern
#1in\vrule}\hrule}}

\def\squiggle#1{\lower1.5ex\hbox{$\sim$}\mkern-14mu #1}

\def\narrower{\advance\leftskip by\parindent \advance\rightskip
by\parindent}

\def\mbox#1#2{\vcenter{\hrule width#1in\hbox{\vrule height#2in
   \hskip#1in\vrule height#2in}\hrule width#1in}}
\def\eqsquare #1:#2:{\vcenter{\hrule width#1\hbox{\vrule height#2
   \hskip#1\vrule height#2}\hrule width#1}}
\def\inbox#1#2#3{\vcenter to #2in{\vfil\hbox to
#1in{$$\hfil#3\hfil$$}\vfil}}
\def\strutdepth{\dp\strutbox}
\def\marbul{\strut\vadjust{\kern-\strutdepth\specialbul}}
\def\specialbul{\vtop to \strutdepth{
    \baselineskip\strutdepth\vss\llap{$\bullet$\qquad}\null}}
\def\Bcomma{\lower6pt\hbox{$,$}}    
\def\bcomma{\lower3pt\hbox{$,$}}    

\def\sl{\scrsf}

\def\updots{\mathinner{\mskip 1mu\raise 1pt\hbox{.}
    \mskip 2mu\raise 4pt\hbox{.}\mskip 2mu
    \raise 7pt\vbox{\kern 7pt\hbox{.}}\mskip 1mu}}

\def\square{\kern1pt\vbox{\hrule height 1.2pt\hbox{\vrule width
1.2pt\hskip 3pt
   \vbox{\vskip 6pt}\hskip 3pt\vrule width 0.6pt}\hrule height
0.6pt}\kern1pt}
\def\ssquare{\kern1pt\vbox{\hrule height .6pt\hbox{\vrule width
.6pt\hskip 3pt
   \vbox{\vskip 6pt}\hskip 3pt\vrule width 0.6pt}\hrule height
0.6pt}\kern1pt}
\def\lege{\hbox{$ {     \lower.40ex\hbox{$>$}
                   \atop \raise.20ex\hbox{$<$}
                   }     $}  }

\def\rege{\hbox{$ {     \lower.40ex\hbox{$<$}
                   \atop \raise.20ex\hbox{$>$}
                   }     $}  }

\def\lapp{\hbox{$ {     \lower.40ex\hbox{$<$}
                   \atop \raise.20ex\hbox{$\sim$}
                   }     $}  }
\def\rapp{\hbox{$ {     \lower.40ex\hbox{$>$}
                   \atop \raise.20ex\hbox{$\sim$}
                   }     $}  }

\def\tridots{\hbox{$ {     \lower.40ex\hbox{$.$}
                   \atop \raise.20ex\hbox{$.\,.$}
                   }     $}  }
\def\Times{\times\hskip-2.3pt{\raise.25ex\hbox{{$\scriptscriptstyle|$}}}}

\def\rightonleft{\hbox{$ {     \lower.40ex\hbox{$\longrightarrow$}
                   \atop \raise.20ex\hbox{$\longleftarrow$}
                   }     $}  }

\def\pmb#1{\setbox0=\hbox{#1}%
\kern-.025em\copy0\kern-\wd0
\kern.05em\copy0\kern-\wd0
\kern-.025em\raise.0433em\box0 }
%
%
\font\fivebf=cmbx5
\font\sixbf=cmbx6
\font\sevenbf=cmbx7
\font\eightbf=cmbx8
\font\ninebf=cmbx9
\font\tenbf=cmbx10

\font\bfmone=cmbx10 scaled\magstep1

\font\sevenit=cmti7
\font\eightit=cmti8
\font\nineit=cmti9
\font\tenit=cmti10

\font\itmone=cmti10 scaled\magstep1

\font\fiverm=cmr5
\font\sixrm=cmr6
\font\sevenrm=cmr7
\font\eightrm=cmr8
\font\ninerm=cmr9
\font\tenrm=cmr10

\font\rmmone=cmr10 scaled\magstep1

\def\fontone{\def\rm{\fcm0\rmmone}%
  \textfont0=\rmmone \scriptfont0=\tenrm \scriptscriptfont0=\sevenrm
  \textfont1=\itmone \scriptfont1=\tenit \scriptscriptfont1=\sevenit
  \def\it{\fcm\itfcm\itmone}%
  \textfont\itfcm=\itmone
  \def\bf{\fcm\bffcm\bfmone}%
  \textfont\bffcm=\bfmone \scriptfont\bffcm=\tenbf
   \scriptscriptfont\bffcm=\sevenbf
  \tt \ttglue=.5em plus.25em minus.15em
  \normalbaselineskip=25pt
  \let\sc=\tenrm
  \let\big=\tenbig
  \setbox\strutbox=\hbox{\vrule height10.2pt depth4.2pt width\z@}%
  \normalbaselines\rm}



\font\ninerm=cmr9
\font\eightrm=cmr8
\font\sixrm=cmr6

\font\ninei=cmmi9
\font\eighti=cmmi8
\font\sixi=cmmi6
\skewchar\ninei='177 \skewchar\eighti='177 \skewchar\sixi='177

\font\ninesy=cmsy9
\font\eightsy=cmsy8
\font\sixsy=cmsy6
\skewchar\ninesy='60 \skewchar\eightsy='60 \skewchar\sixsy='60

\font\ninebf=cmbx9
\font\eightbf=cmbx8
\font\sixbf=cmbx6

\font\ninett=cmtt9
\font\eighttt=cmtt8

\hyphenchar\tentt=-1 
\hyphenchar\ninett=-1
\hyphenchar\eighttt=-1

\font\ninesl=cmsl9
\font\eightsl=cmsl8

\font\nineit=cmti9
\font\eightit=cmti8


\newskip\ttglue
\def\tenpoint{\def\rm{\fcm0\tenrm}%
  \textfont0=\tenrm \scriptfont0=\sevenrm \scriptscriptfont0=\fiverm
  \textfont1=\teni \scriptfont1=\seveni \scriptscriptfont1=\fivei
  \textfont2=\tensy \scriptfont2=\sevensy \scriptscriptfont2=\fivesy
  \textfont3=\tenex \scriptfont3=\tenex \scriptscriptfont3=\tenex
  \def\it{\fcm\itfcm\tenit}%
  \textfont\itfcm=\tenit
  \def\sl{\fcm\slfcm\tensl}%
  \textfont\slfcm=\tensl
  \def\bf{\fcm\bffcm\tenbf}%
  \textfont\bffcm=\tenbf \scriptfont\bffcm=\sevenbf
   \scriptscriptfont\bffcm=\fivebf
  \def\tt{\fcm\ttfcm\tentt}%
  \textfont\ttfcm=\tentt
  \tt \ttglue=.5em plus.25em minus.15em
  \normalbaselineskip=16pt
  \let\sc=\eightrm
  \let\big=\tenbig
  \setbox\strutbox=\hbox{\vrule height8.5pt depth3.5pt width\z@}%
  \normalbaselines\rm}

\def\ninepoint{\def\rm{\fcm0\ninerm}%
  \textfont0=\ninerm \scriptfont0=\sixrm \scriptscriptfont0=\fiverm
  \textfont1=\ninei \scriptfont1=\sixi \scriptscriptfont1=\fivei
  \textfont2=\ninesy \scriptfont2=\sixsy \scriptscriptfont2=\fivesy
  \textfont3=\tenex \scriptfont3=\tenex \scriptscriptfont3=\tenex
  \def\it{\fcm\itfcm\nineit}%
  \textfont\itfcm=\nineit
  \def\sl{\fcm\slfcm\ninesl}%
  \textfont\slfcm=\ninesl
  \def\bf{\fcm\bffcm\ninebf}%
  \textfont\bffcm=\ninebf \scriptfont\bffcm=\sixbf
   \scriptscriptfont\bffcm=\fivebf
  \def\tt{\fcm\ttfcm\ninett}%
  \textfont\ttfcm=\ninett
  \tt \ttglue=.5em plus.25em minus.15em
  \normalbaselineskip=11pt
  \let\sc=\sevenrm
  \let\big=\ninebig
  \setbox\strutbox=\hbox{\vrule height8pt depth3pt width\z@}%
  \normalbaselines\rm}

\def\eightpoint{\def\rm{\fcm0\eightrm}%
  \textfont0=\eightrm \scriptfont0=\sixrm \scriptscriptfont0=\fiverm
  \textfont1=\eighti \scriptfont1=\sixi \scriptscriptfont1=\fivei
  \textfont2=\eightsy \scriptfont2=\sixsy \scriptscriptfont2=\fivesy
  \textfont3=\tenex \scriptfont3=\tenex \scriptscriptfont3=\tenex
  \def\it{\fcm\itfcm\eightit}%
  \textfont\itfcm=\eightit
  \def\sl{\fcm\slfcm\eightsl}%
  \textfont\slfcm=\eightsl
  \def\bf{\fcm\bffcm\eightbf}%
  \textfont\bffcm=\eightbf \scriptfont\bffcm=\sixbf
   \scriptscriptfont\bffcm=\fivebf
  \def\tt{\fcm\ttfcm\eighttt}%
  \textfont\ttfcm=\eighttt
  \tt \ttglue=.5em plus.25em minus.15em
  \normalbaselineskip=9pt
  \let\sc=\sixrm
  \let\big=\eightbig
  \setbox\strutbox=\hbox{\vrule height7pt depth2pt width\z@}%
  \normalbaselines\rm}



\magnification=1200
\newbox\leftpage
\newdimen\fullhsize
\newdimen\hstitle
\newdimen\hsbody
\tolerance=600\hfuzz=2pt
\hoffset=0.1truein \voffset=0.1truein
\hsbody=\hsize \hstitle=\hsize
\font\titlefont=cmr10 scaled\magstep3
\font\secfont=cmbx10 scaled\magstep1

\def\nolabels{\def\eqnlabel##1{}\def\eqlabel##1{}\def\reflabel##1{}}
\def\writelabels{\def\eqnlabel##1{%
{\escapechar=` \hfill\rlap{\hskip.09in\string##1}}}%
\def\eqlabel##1{{\escapechar=` \rlap{\hskip.09in\string##1}}}%
\def\reflabel##1{\noexpand\llap{\string\string\string##1\hskip.31in}}}
\nolabels
\def\title#1{ \nopagenumbers\hsize=\hsbody%
\centerline{ {\titlefont #1} }%
\pageno=0}
\def\author#1{\vskip 1 truecm%
\centerline{{\sl #1}}%
\centerline{Center for Theoretical Physics}%
\centerline{Laboratory for Nuclear Science}%
\centerline{and Department of Physics}%
\centerline{Massachusetts Institute of Technology}%
\centerline{Cambridge, Massachusetts 02139}}
\def\abstract#1{\centerline{\bf ABSTRACT}\nobreak\medskip\nobreak\par #1}

\global\newcount\secno \global\secno=0
\global\newcount\meqno \global\meqno=1
\def\newsec#1{\global\advance\secno by1
\xdef\secsym{\ifcase\secno
\or I\or II\or III\or IV\or V\or VI\or VII\or VIII\or IX\or X\fi }
\global\meqno=1
\bigbreak\bigskip
\noindent{\secfont #1}\par\nobreak\medskip\nobreak}
\xdef\secsym{}


\def\eqnn#1{\xdef #1{(\the\meqno)}%
\global\advance\meqno by1\eqnlabel#1}
\def\eqna#1{\xdef #1##1{\hbox{$(\the\meqno##1)$}}%
\global\advance\meqno by1\eqnlabel{#1$\{\}$}}
\def\eqn#1#2{\xdef #1{(\the\meqno)}\global\advance\meqno by1%
$$#2\eqno#1\eqlabel#1$$}
%
%
%

\def\eqnla#1#2#3{\xdef #2{(\the\meqno a)}
\xdef #1{(\the\meqno)}
$$#3\eqno#2\eqlabel#1\eqlabel#2$$}
\def\eqnlb#1#2{\xdef #1{(\the\meqno b)}
$$#2\eqno#1\eqlabel#1$$}
\def\eqnlc#1#2{\xdef #1{(\the\meqno c)}
$$#2\eqno#1\eqlabel#1$$}
\def\eqnld#1#2{\xdef #1{(\the\meqno d)}
$$#2\eqno#1\eqlabel#1$$}
\def\eqnle#1#2{\xdef #1{(\the\meqno e)}
$$#2\eqno#1\eqlabel#1$$}
\def\eqnlf#1#2{\xdef #1{(\the\meqno f)}
$$#2\eqno#1\eqlabel#1$$}
\def\eqnlbend#1#2{\xdef #1{(\the\meqno b)}\global\advance\meqno by1%
$$#2\eqno#1\eqlabel#1$$}
\def\eqnlcend#1#2{\xdef #1{(\the\meqno c)}\global\advance\meqno by1%
$$#2\eqno#1\eqlabel#1$$}
\def\eqnldend#1#2{\xdef #1{(\the\meqno d)}\global\advance\meqno by1%
$$#2\eqno#1\eqlabel#1$$}
\def\eqnleend#1#2{\xdef #1{(\the\meqno e)}\global\advance\meqno by1%
$$#2\eqno#1\eqlabel#1$$}
\def\eqnlfend#1#2{\xdef #1{(\the\meqno f)}\global\advance\meqno by1%
$$#2\eqno#1\eqlabel#1$$}
\def\eqnlgend#1#2{\xdef #1{(\the\meqno g)}\global\advance\meqno by1%
$$#2\eqno#1\eqlabel#1$$}
\global\newcount\ftno \global\ftno=1
\def\foot#1{{\baselineskip=12pt plus 1pt\footnote{$^{\the\ftno}$}{#1}}%
\global\advance\ftno by1}
\global\newcount\refno \global\refno=1
\newwrite\rfile
\def\ref#1#2{\the\refno\nref#1{#2}}
\def\nref#1#2{\xdef#1{\the\refno}%
\ifnum\refno=1\immediate\openout\rfile=refs.tmp\fi%
\immediate\write\rfile{\noexpand\item{#1\ }\reflabel{#1}#2}%
\global\advance\refno by1}
\def\addref#1{\immediate\write\rfile{\noexpand\item{}#1}}

\def\refnn#1#2{\nref#1{#2}}
\def\nref#1#2{\xdef#1{\the\refno}%
\ifnum\refno=1\immediate\openout\rfile=refs.tmp\fi%
\immediate\write\rfile{\noexpand\item{#1\ }\reflabel{#1}#2}%
\global\advance\refno by1}
\def\addref#1{\immediate\write\rfile{\noexpand\item{}#1}}



\def\frac#1#2{{#1\over#2}}

\baselineskip=24pt

\def\s{\space}

\def\n{\nu}

\def\rpln{r^{|l+\n|}}

\def\u{\Psi}

\def\wd{$\hat{\Psi}$}

\pageno=0

\centerline{{\bf PERTURBATIVE BOSONIC END ANYON SPECTRA }}
\centerline{{\bf AND CONTACT INTERACTIONS}\footnote{*}
{This work is
supported in part by funds provided by the U.S. Department of Energy (D.O.E.)
under contract \#DE-AC02-76ER03069.}}
\vskip 50pt
\bigskip
\centerline{G. Amelino-Camelia}
\vskip 18pt
\baselineskip 12pt plus 0.2pt minus 0.2pt
\centerline{{\it Center for Theoretical Physics}}
\centerline{{\it Laboratory for Nuclear Science}}
\centerline{{\it and Department of Physics}}
\centerline{{\it Massachusetts Institute of Technology}}
\centerline{{\it Cambridge, Massachusetts 02139 USA}}
\vskip 2.2cm
\centerline{\bf ABSTRACT }
Bosonic end perturbative calculations for quantum mechanical
anyon systems require a regularization.
I regularize by
adding a specific $\delta$-function potential to the Hamiltonian.
The reliability of this regularization procedure is verified by
comparing its results for the 2-anyon in harmonic potential system with
the known exact solutions.
I then use the $\delta$-function regularized bosonic end perturbation theory
to test some recent conjectures concerning the unknown
portion of the many-anyon spectra.

\vskip 1.5cm
\centerline{Submitted to: Physics Letters B}
\vfill
\noindent{MIT-CTP-2242 \hfill September, 1993}
\eject
\vfill
\baselineskip 24pt plus 0.2pt minus 0.2pt

In 2+1 dimensions the triviality of the rotation group SO(2) allows the
existence of anyons:
particles with
``anomalous"
spin and statistics$^{[\ref\wil{J. M. Leinaas and J. Myrheim,
Nuovo Cimento B 37 (1977) 1;
G. A. Goldin, R. Menikoff, and D. H. Sharp, J. Math.
Phys.  21 (1980) 650;  22 (1981) 1664;
F. Wilczek,  Phys. Rev. Lett. 48 (1982) 1144; 49 (1982) 957 .
Also see: F. Wilczek, Fractional Statistics and Anyon Superconductivity,
(World Scientific, 1990).
}]}$.
It has been
conjectured$^{[\ref\lh{
R. B. Laughlin, Phys. Rev. Lett. 50 (1983) 1395;
B. I. Halperin, Phys. Rev. Lett. 52 (1984) 1583.}]}$
that anyons
might play an important role in
the understanding of some planar condensed matter phenomena, most notably
the fractional quantum Hall effect, and
this has motivated numerous recent investigations of quantum mechanical
anyon systems
and of field theory realizations of anomalous
quantum statistics.

One would like to develop a description of anyons as complete and
intuitive as the ones available for bosons and fermions, but thus far
this program has had only partial success.
In particular, the anyon quantum mechanics
is not completely understood.
Even for the simple cases of $N$ identical anyons
in an external magnetic field and/or harmonic potential
the complete set of eigensolutions
is known$^{[\wil]}$ only for $N=2$.
For systems of more than 2 anyons
just a particular
class of eigensolutions has been
found$^{[\ref\tru{Y. S. Wu, Phys. Rev. Lett. 53 (1984) 111;
G. V. Dunne, A. Lerda, and C. A. Trugenberger,
Mod. Phys. Lett. A 6 (1991) 2819; G. V. Dunne, A. Lerda,
S. Sciuto, and C. A. Trugenberger, Nucl. Phys. B 370 (1992) 601;
K. H. Cho and C. Rim, Ann. Phys. 213 (1992) 295;
Chihong Chou,
Phys. Lett. A 155 (1991) 245;
S. Mashkevich, Int. J. Mod. Phys. A 7 (1992) 7931.},
\ref\cho{C. Chou, Phys. Rev. D 44 (1991) 2533, D 45 (1992) 1433(E).}]}$.
In order to obtain some information concerning the ``missing eigensolutions"
various approximation techniques have been used:
``bosonic end perturbation theory"$^{[\cho-
\refnn\comtet{A. Comtet, J. McCabe, and S. Ouvry, Phys. Lett. B 260
(1991) 372;
J. McCabe and S. Ouvry, Phys. Lett. B 260 (1991) 113;
A. Dasnieres de Veigy and S. Ouvry, Phys. Lett. B 291 (1992) 130;
Nucl. Phys. B 388 (1992) 715.}
\refnn\sen{D. Sen, Nucl. Phys. B 360 (1991) 397.}
\refnn\gac{G. Amelino-Camelia, Phys. Lett. B 286 (1992) 97.}
\refnn\mit{C. Chou, L. Hua,
and G. Amelino-Camelia, Phys. Lett. 286 (1992) 329.}
\ref\papsix{G. Amelino-Camelia, Phys. Lett. 299 (1992) 83.}]}$,
``fermionic end perturbation theory"$^{[\cho,
\ref\verbapert{M. Sporre,
J. J. M. Verbaarschot, and I. Zahed, Nucl. Phys. B 389 (1993) 645.}]}$,
and numerical methods$^{[\ref\verba{M. Sporre,
J. J. M. Verbaarschot, and I. Zahed, Phys. Rev. Lett.67 (1991) 1813;
SUNY preprint SUNY-NTG-91/40, October 1991.},
\ref\murthy{M. V. N. Murthy, J. Law, M. Brack, and R. K.
Bhaduri, Phys. Rev. Lett. 67 (1991) 1817.}]}$.

The results presented in this letter are in the framework of the
bosonic end perturbation theory. This technique concerns the study
of anyons with small statistical parameter $\nu$ (here
defined following the
convention of Refs.[\gac,\papsix]), by using a perturbative expansion
in $\nu$.
The unperturbed wave functions are bosonic\footnote{*}
{In fermionic end perturbation theory $\nu \simeq 1$
and the unperturbed wave functions are fermionic.}; in fact in the
limit $\nu \rightarrow 0$ anyons acquire bosonic statistics$^{[\wil]}$.
Due to the known$^{[\wil,\gac]}$ non-analyticity of the
limit $\nu \rightarrow 0$, in applying this perturbative approach
some divergencies appear, and therefore regularization procedures are
necessary$^{[\cho-\papsix]}$.
For example, the magnetic
gauge$^{[\wil,\gac]}$ Hamiltonian which describes the relative
motion\footnote{*}
{The center of mass motion is simply a free motion and
it is essentially irrelevant for the discussion in this letter.}
of 2 anyons in a common harmonic well is given by
\eqn\ham{H_2 = - {1 \over r} \partial_{r} (r \partial_{r})
- {1 \over r^2}  \partial_{\phi}^{2} + r^2
 -{ 2i\n \over  r^2}\partial_{\phi} + { \nu^2 \over r^2} ~,}
and logarithmic divergencies in the bosonic end perturbative analysis
originate from
some of the matrix elements of the $\nu^2 / r^2$ term, like
\eqn\div{<\Omega_0| {\n^2 \over r^2} |\Omega_0>
= 2 \n^2 \int^{\infty}_{0} dr ~ {\exp( -r^2 ) \over r} ~\sim \infty
{}~,}
where $|\Omega_0> \equiv ({e^{ - {r^2  / 2}} )/ \pi^{1 / 2}}$ is the
unperturbed (i.e. bosonic) ground state.

Many of the results presently available on
the unknown portion of the anyon spectra,
have been obtained using
bosonic end perturbation theory in the analysis of few anyon systems.
However,
as already emphasized in Ref.[\ref\lozanob{
O. Bergman and G. Lozano, MIT preprint MIT-CTP-2182, January 1993.}],
the regularization
procedures used in these calculations
require rather arbitrary manipulations.

In this letter I calculate for arbitrary $N$ and
to second order in $\nu$
some of the eigenenergies of the $N$-anyon
in harmonic potential system.
This analysis extends some of the results obtained in the literature
on the anyon spectra,
and could also be useful in the search of new exact solutions to the
$N$-anyon problem by leading to some ``educated {\it ansaetze}".
Moreover, in my calculations I use a new regularization procedure,
based on the introduction of a repulsive $\delta$-function
potential, which has a simpler physical interpretation than
the ones used in the literature.

Let me start by observing that,
because the anyonic wave functions vanish\footnote{*}
{The possibility of anyons with wave functions which do not vanish at
the points of overlap has also been considered in the literature
(for example this subject is discussed in
Ref.[${\ref\bourdeau{M. Bourdeau and R.D. Sorkin,
Phys. Rev. D45 (1992) 687.}}$]), but
in this letter only
the conventional ``non-colliding" anyons are considered.}
at the points of overlap,
the addition of a repulsive
$\delta$-function potential to the Hamiltonian $H_N$
of a quantum mechanical $N$-anyon system has no physical consequences
(i.e. the exact eigensolutions are unaffected by it),
but it can be used to implement in the bosonic end
perturbation theory the hard core boundary
condition$^{[\ref\gacnext{More on the physical interpretation of the
$\delta$-function regularization, together with a detailed comparison
to the regularization procedures previously used in the literature,
will be included in a forthcoming longer pubblication:
G. Amelino-Camelia, in preparation.}]}$.
I can therefore apply small-$\nu$ perturbation theory,
rather than to the original Hamiltonian $H_N$,
to the equivalent
Hamiltonian $H^{\delta}_N$, given by\footnote{**}
{Note that in Ref.[${\ref\kogan{I.I. Kogan, Phys.
Lett. B 262 (1991) 83;
I.I. Kogan and G.W. Semenoff, Nucl. Phys. B 368 (1992) 718.}}$]
an Hamiltonian of the type $H^{\delta}_N$ was already
considered, as a result
of an analysis of an induced anyon magnetic moment.}
\eqn\hndelgen{\eqalign{ H_N^{\delta} \equiv
& H_{N} +
2 \pi |\nu| \sum_{m < n} \delta^{(2)}({\bf r_{mn}})
{}~,}}
where ${\bf r_{mn}}$ is the relative position of the m-th and n-th anyon
in the system.

In the following we will see that the $\delta$-function potential added
in \hndelgen\s eliminates the divergencies of bosonic end perturbation
theory.
This is not surprising in light of the results obtained
in some recent investigations
of field theory realizations of fractional
quantum statistics
in which the addition of a quartic contact term,
the field theory analog of a quantum mechanical
$\delta$-function potential,
has been shown
to eliminate some divergencies$^{[\lozanob,
\ref\lozanoa{G. Lozano, Phys. Lett. B 283 (1992) 70.},
\ref\valle{M.A. Valle Basagoiti, Phys. Lett. B306 (1993) 307;
R. Emparan and M.A. Valle Basagoiti, Universidad del Pais Vasco
preprint EHU-FT-93-5, April 1993.}]}$.

As a first test of the reliability of
this ``$\delta$-function regularization", it is convenient to discuss
its application to the 2-anyon in harmonic potential problem;
the regularized Hamiltonian is
\eqn\hamdel{
H^\delta_2 \equiv
- {1 \over r} \partial_{r} (r \partial_{r})
- {1 \over r^2}  \partial_{\phi}^{2} + r^2
-{ 2i\n \over  r^2}\partial_{\phi}
 + 2 \pi |\nu| \delta^{(2)}({\bf r})
+ { \nu^2 \over r^2}
=H_2 + 2 \pi |\nu| \delta^{(2)}({\bf r})
{}~.}
Although the $H^\delta_2$-eigenproblem is equivalent to the
$H_2$-eigenproblem, $H^\delta_2$ is more suitable for perturbation theory;
in fact, the added $\delta$-function potential leads to
divergencies which exactly cancel those introduced by the $\nu^2 / r^2$ term,
rendering finite the results of bosonic end perturbation theory.
Let me illustrate this mechanism by verifying that
the first and second order eigenenergies and
the first order eigenfunctions obtained with
the $\delta$-function regularized
bosonic end perturbation theory are
finite and
in agreement with the known eigensolutions
of the 2-anyon in harmonic potential problem, which are given
by$^{[\wil]}$

\baselineskip 12pt plus 0.2pt minus 0.2pt
\eqnla\sol\solb{
E^{exact}_{n,l,\nu} = ( 4n + 2|l+\n| + 2)
{}~,}

\eqnlbend\sola{
|\u_{n,l,\nu}^{exact}> =
N_{n,l}^{\nu}~ \rpln ~
e^{ - {r^2 \over 2} + i l \phi }~
L_{n}^{|l+\nu|}(r^2)
{}~,}
\smallskip
\baselineskip 24pt plus 0.2pt minus 0.2pt

\noindent
where the $L_{n}^{x}$ are
Laguerre polynomials, and the $N_{n,l}^{\nu}$ are normalization
constants.

\noindent
My analysis is limited to the states with $l=0$.
For the states with $l \ne 0$ no divergence is present to begin
with$^{[\cho-\gac]}$,
and the consistency of the $\delta$-function regularization
can be verified in complete analogy with the corresponding results
obtained for the other regularization procedures. [N.B. The
$\delta$-function potential does not contribute to the matrix elements
involving unperturbed
states with $l \ne 0$, because these states vanish
for ${\bf r} = 0$.]

Concerning the first order energies,
one easily finds
\eqn\eonecoppia{
E^{(1)}_{n,0,\nu} = <\u^{(0)}_{n,0}|
-{ 2i\n \over  r^2}\partial_{\phi}
 + 2 \pi |\nu| \delta^{(2)}({\bf r})
|\u^{(0)}_{n,0}> = 2 |\nu|
{}~,}
where the unperturbed eigenfunctions are
\eqn\unpertcoppia{
|\u^{(0)}_{n,l}> \equiv
({n! \over {\pi (n+l)!}})^{1 \over 2}~
r^{|l|} ~
e^{ - {r^2 \over 2} + i l \phi }~
L_{n}^{|l|}(r^2) = |\u_{n,l,0}^{exact}>
{}~}
(i.e. the $|\u^{(0)}_{n,l}>$ are the
eigenfunctions of the 2-boson in harmonic potential problem).

\noindent
The result \eonecoppia\s is clearly in agreement with Eq.\solb.

The first order eigenfunctions are given by
\eqn\psionecoppia{\eqalign{
|\Psi^{(1)}_{n,0,\nu}> = &
\sum_{m,l \ne n,0}
{<\Psi^{(0)}_{m,l}| ~
-{ 2i\n \over  r^2}\partial_{\phi}
 + 2 \pi |\nu| \delta^{(2)}({\bf r}) ~ |\Psi^{(0)}_{n,0}>
\over E^{(0)}_{n,0}-E^{(0)}_{m,l} } |\Psi^{(0)}_{m,l}>
\cr
= & - {|\n| \over 2\sqrt{\pi }} \sum_{m \ne n}
{L_{m}^{0}(r^2) \over m-n} e^{-{r^2 \over 2}}
{}~.}}
Using properties of
the Laguerre polynomials one can verify (with some algebra) that the
result \psionecoppia\s is in agreement with Eq.\sola. For example
one obtains ($\gamma$ is the Euler constant)
\eqn\psioneto{
|\Psi^{(1)}_{2,0,\nu}> =
 {|\n| \over \sqrt{\pi }}  e^{-{r^2 \over 2}}
\biggl[ {3 \over 2} - r^2 + {1 \over 4} (2\gamma - 3 + 4 ln(r)) L_{2}^{0}(r^2)
\biggr]
{}~,}
which is in perfect agreement with the first order term in the expansion
in $\nu$ of $|\Psi^{exact}_{2,0,\nu}>$.

{}From \hamdel\s one sees that the second order energies
are given by:
\eqn\etwocoppia{
E^{(2)}_{n,0,\nu} ~ = ~
<\u^{(0)}_{n,0}| {\nu^2 \over r^2}
|\u^{(0)}_{n,0}> ~ + ~
<\u^{(0)}_{n,0}| -{ 2i\n \over  r^2}\partial_{\phi}
 + 2 \pi |\nu| \delta^{(2)}({\bf r})
|\u^{(1)}_{n,0,\nu}> ~ \equiv ~ E^{(2,a)}_{n,0,\nu} ~
+ ~ E^{(2,b)}_{n,0,\nu}
{}~.}
Both $E^{(2,a)}_{n,0,\nu}$ and $E^{(2,b)}_{n,0,\nu}$ are divergent, but
the final result is finite. In order to illustrate the details of the
cancellation of the infinities, it is useful to follow a definite
calculation; for example for $E^{(2)}_{2,0,\nu}$ the contributions are:

\baselineskip 12pt plus 0.2pt minus 0.2pt
\eqnla\etworeg\etworega{\eqalign{
E^{(2,a)}_{2,0,\nu} ~ =& ~ <\u^{(0)}_{2,0}| ~ {\nu^2 \over r^2} ~
|\u^{(0)}_{2,0}> ~ = ~
\nu^2 \int^{\infty}_{0} \int^{2 \pi}_{0} dr~ d\phi ~
{\exp( -r^2 ) \over \pi r}~ [L_{2}^{0}(r^2)]^2
\cr
=& ~  \lim_{\epsilon \rightarrow 0}
\int^{\infty}_{\epsilon}  dr ~ \nu^2 ~
{\exp( -r^2 ) \over r}~ [L_{2}^{0}(r^2)]^2 ~ = ~
\nu^2 ~ \lim_{\epsilon \rightarrow 0}
\biggl[-2 \ln(\epsilon) -\gamma -{3 \over 2}\biggr]
{}~,}}

\eqnlbend\etworegb{\eqalign{
E^{(2,b)}_{2,0,\nu} ~ =& ~ <\u^{(0)}_{2,0}| ~
-{ 2i\n \over r^2}\partial_{\phi}
 + 2 \pi |\nu| \delta^{(2)}({\bf r})
{}~ |\u^{(1)}_{2,0,\nu}> ~
= ~ <\u^{(0)}_{2,0}| ~ 2 \pi |\nu| \delta^{(2)}({\bf r}) ~
|\u^{(1)}_{2,0,\nu}>
\cr
=& ~
2 \nu^2 \int^{\infty}_{-\infty} \int^{\infty}_{-\infty}  dr_x ~ dr_y ~
\delta^{(2)}({\bf r}) ~ e^{-r^2} ~ L_{2}^{0}(r^2) ~
\biggl[ {3 \over 2} - r^2 + {1 \over 4} (2\gamma - 3 + 4 ln(r)) L_{2}^{0}(r^2)
\biggr]
\cr
=& ~ \nu^2 ~ \lim_{\epsilon \rightarrow 0}
\biggl[2 ln(\epsilon) + {3 \over 2} + \gamma \biggr]
{}~.}}
\smallskip
\baselineskip 24pt plus 0.2pt minus 0.2pt

\noindent
Note that,
in order to see the cancellation of infinities and evaluate the left-over
finite result,
I introduced a cut-off $\epsilon$, which will be ultimately removed by
taking the limit $\epsilon\rightarrow 0$.
In general a similar cut-off must be introduced in all the divergent
matrix elements
of $r^{-2}$ and $\delta^{(2)}({\bf r})$
by using

\baselineskip 12pt plus 0.2pt minus 0.2pt
\eqnla\reg\rega{
\int^{\infty}_{-\infty} \int^{\infty}_{-\infty}  dr_x ~ dr_y ~
{1 \over r^2}~f(r_x,r_y) =
\lim_{\epsilon \rightarrow 0}
\int^{\infty}_{\epsilon} \int^{2 \pi}_{0} r ~ dr ~ d\phi ~
{1 \over r^2}~f(r \cos \phi,r \sin \phi)
{}~,}

\eqnlbend\regb{
\int^{\infty}_{-\infty} \int^{\infty}_{-\infty}  dr_x ~ dr_y ~
\delta^{(2)}({\bf r})~f(r_x,r_y) =
\lim_{\epsilon \rightarrow 0}
f({\epsilon \over \sqrt{2}},{\epsilon \over \sqrt{2}})
{}~.}
\smallskip
\baselineskip 24pt plus 0.2pt minus 0.2pt

{}From Eqs.\etwocoppia\s and \etworeg\s one concludes
that $E^{(2)}_{2,0,\nu}=0$, and this is in agreement with Eq.\solb,
which indicates that the 2-anyon in harmonic potential spectra
is linear in $\nu$.

This completes my test of
the reliability of the $\delta$-function regularization.
I am now ready to calculate perturbatively from the bosonic end to
second order in the statistical parameter $\nu$\space and for
arbitrary $N$
the eigenenergies of some $N$-anyon in harmonic potential states.

The $\delta$-function regularized Hamiltonian which
describes the relative motion
of $N$ identical anyons in an harmonic potential
is given by
\eqn\hamcalc{H_N^{\delta}=H^{(0)} + H_L^{(1)} + H_\delta^{(1)} + H^{(2)} }
where $H_0$ is the relative motion Hamiltonian for $N$
bosons in an harmonic potential, and

\baselineskip 12pt plus 0.2pt minus 0.2pt
\eqnla\hpert\hperta{
H_L^{(1)} \equiv
{\nu \over 2} \sum_{m \ne n}
{1 \over  |z_n - z_m|^2} L_{n,m}
{}~,}
\eqnlb\hpertb{
H_\delta^{(1)} \equiv
2 \pi |\nu| \sum_{m < n} \delta^{(2)}(z_m - z_n )
{}~,}
\eqnlcend\hpertc{
H_2={\nu^2 \over 4} \sum_{m\ne n, n \ne k }
\left({1 \over (z_n-z_m)(z_n^*-z_k^*)} +~ h.c. ~\right)
{}~.}
\smallskip
\baselineskip 24pt plus 0.2pt minus 0.2pt

\noindent
The operators $L_{n,m}$ are given by
\eqn\hamb{\eqalign{&
L_{n,m} \equiv  (z_n - z_m)
\left({\partial \over \partial z_n} -{\partial \over \partial z_m}\right)
- (z_n^* - z_m^*)
\left({\partial \over \partial z_n^*} -
{\partial \over \partial z_m^*}\right) ~,} }
and I am using the conventional notation $z_n \equiv x_n + i y_n$,
$z_n^* \equiv x_n - i y_n$.

The three $N$-anyon states whose energies I shall evaluate perturbatively
are the ones which correspond, in
the limit $\nu \rightarrow 0$,
to the following bosonic states

\baselineskip 12pt plus 0.2pt minus 0.2pt
\eqnla\state\statea{
|\Omega> \equiv {1 \over {\sqrt {\pi^{N-1}}}} ~
exp \left[ \sum_{n=1}^{N-1} |u_n(\{ z_i \})|^2 \right]
{}~,}
\eqnlb\stateb{
|+> \equiv {1 \over \sqrt{2 ~ (N-1) ~ \pi^{N-1}}}
\left( \sum_{n=1}^{N-1} u^2_n(\{ z_i \}) \right)
exp \left[ \sum_{n=1}^{N-1} |u_n(\{ z_i \})|^2 \right]
{}~,}

\eqnlcend\statec{
|-> \equiv |+>^*  ~,}
\smallskip
\baselineskip 24pt plus 0.2pt minus 0.2pt

\noindent
where $u_n(\{ z_i \}) \equiv (z_{1}+z_{2}+ .... + z_n
- n ~ z_{n+1})/ \sqrt{n(n+1)}$.

\noindent
$|\Omega>$ is the $N$-boson groundstate (and the corresponding
perturbative results give the approximate $N$-anyon groundstate energy
for small $\nu$), and the states $|\pm>$ are in the first excited
bosonic energy level and have angular momentum $\pm 2$.

\noindent
It is easy to verify that $E^{(1,2)}_\Omega (\nu)=E^{(1,2)}_\Omega (-\nu)$,
$E^{(1)}_+ (\nu)=-E^{(1)}_- (-\nu)$, and
$E^{(2)}_+ (\nu)=E^{(2)}_- (-\nu)$; therefore, I can limit
the calculations to the case $\nu >0$
without any loss of generality.

The first order energies can be easily calculated, they are given by

\baselineskip 12pt plus 0.2pt minus 0.2pt
\eqnla\eonen\eonena{
E^{(1)}_\Omega =
<\Omega| H_L^{(1)} + H_\delta^{(1)}
|\Omega> =
{N (N-1) \over 2} \nu
{}~,}
\eqnlbend\stateb{
E^{(1)}_\pm =
<\pm|
H_L^{(1)} + H_\delta^{(1)}
|\pm> =
{N (N-2) \over 2} \nu \pm {N \over 2} \nu
{}~.}
\smallskip
\baselineskip 24pt plus 0.2pt minus 0.2pt

Concerning the evaluation of the second order energies,
let us start by noticing that from \hamcalc\s and \hpert\s
it follows that

\baselineskip 12pt plus 0.2pt minus 0.2pt
\eqnla\etwopieces\etwopiecesa{
E_{\Psi^{(0)}}^{(2)}= <\Psi^{(0)}| H^{(2)}
|\Psi^{(0)}>
+E_{\Psi^{(0)},L}^{(2)}+E_{\Psi^{(0)},\delta}^{(2)}
{}~,}
\eqnlb\etwopiecesb{E_{\Psi^{(0)},L}^{(2)}
\equiv \sum_{|m> \notin D} { <\Psi^{(0)}|
H_L^{(1)} + H_\delta^{(1)} |m>
<m| H_L^{(1)} |\Psi^{(0)}>
\over {E^{(0)}-E^{(0)}_m}}
{}~,}
\eqnlcend\etwopiecesc{E_{\Psi^{(0)},\delta^{(2)}} \equiv
\sum_{|m> \notin D} { <\Psi^{(0)}|
H_L^{(1)} + H_\delta^{(1)} |m>
<m| H_\delta^{(1)} |\Psi^{(0)}>
\over {E^{(0)}-E^{(0)}_m}}
{}~,}
\smallskip
\baselineskip 24pt plus 0.2pt minus 0.2pt

\noindent
where $D$ is  the space of bosonic states with energy $E^{(0)}$ ($dim D$ is
the ``degeneracy'' of $|\Psi^{(0)}>$).
Using the symmetries of $H^{(2)}$ and of the unperturbed wave
functions \state,
one easily obtains

\baselineskip 12pt plus 0.2pt minus 0.2pt
\eqnla\nusquare\nusquarea{
<\Omega| H^{(2)}
|\Omega> ~ = ~
\lim_{\epsilon \rightarrow 0}
\biggl[ {\nu^2 \over 4}~N(N-1)~
\biggl( 2 (N-2) \ln({4 \over 3}) - \gamma - \ln(\epsilon) \biggr) \biggr]
{}~,}
\eqnlbend\nusquareb{<\pm| H^{(2)}
|\pm> ~ = ~ \lim_{\epsilon \rightarrow 0}
\biggl[ {\nu^2 \over 8} ~N
\biggl( 9 - 4 N + 4 (N+1) (N-2) \ln({4 \over 3})
- 2 (N-2) (\gamma + \ln(\epsilon)) \biggr) \biggr]
{}~,}
\smallskip
\baselineskip 24pt plus 0.2pt minus 0.2pt

\noindent
where I introduced the cut-off $\epsilon$ using \reg.

For the states \state\s the evaluation of $E_{\Psi^{(0)},L}^{(2)}$
and $E_{\Psi^{(0)},\delta}^{(2)}$
(which is usually possible only numerically) is
relevantly simplified by
the following results
\eqn\cidn{ \eqalign{&{ 1\over |z_1-z_2|^2} L_{12} ~ |\pm> ~
= ~ {1\over 4}
\left( [C_{12}, H^{(0)}]- \pi \delta^{(2)}({z_1-z_2 \over {\sqrt 2}})
+ 1 \right)
L_{12} ~ |\pm>  ~, \cr
&<m| ~ \pi ~ \delta^{(2)}({z_1-z_2 \over {\sqrt 2}}) ~ |\Omega (\pm)>
{}~ = ~ <m| ~ [C_{12}, H^{(0)}] ~ |\Omega (\pm)>
{}~, \cr
&H^{(0)} ~ L_{12} ~ |\pm> ~ = ~ E^{(0)}_{\pm} ~ L_{12} ~ |\pm> ~~;
{}~~ L_{12} ~ |\Omega> ~ = ~0
,~ }}
where $C_{mn} \equiv [ln( {|z_m-z_n|^2 \over 2}) + \gamma - 1]/2$.
Using the properties \cidn\s
(and introducing again the cut-off $\epsilon$ using \reg)
one finds

\baselineskip 12pt plus 0.2pt minus 0.2pt
\eqnla\rescomp\rescompa{
E_{\Omega,L}^{(2)}=0
{}~,}
\eqnlb\rescompb{\eqalign{
E_{\Omega,\delta}^{(2)} =
\lim_{\epsilon \rightarrow 0}
\biggl[ {\nu^2 \over 4}~N(N-1)~
\biggl( \ln(\epsilon) + \gamma
- 2 (N-2) \ln({4 \over 3}) \biggr) \biggr]
{}~,}}
\eqnlc\rescompc{\eqalign{
E_{\pm,L}^{(2)} =
\pm {3 \over 8} \nu^2
N (N-2) \ln({3 \over 4})
+ {\nu^2 \over 16} ~N
\biggl(   5 N - 12 - 18 (N-2) \ln({4 \over 3})   \biggr)
{}~,}}
\eqnldend\rescompd{\eqalign{
E_{\pm,\delta}^{(2)} =
& \pm {3 \over 8} \nu^2 N (N-2) \ln({3 \over 4})
+ {\nu^2 \over 16} ~N~\lim_{\epsilon \rightarrow 0}
\biggl[   3 N - 6
+ 4 (N-2) (\gamma + \ln(\epsilon))
\cr
&~~~~~~~~~~~~~~~~~~~~~~~~~~
{}~~~~~~~~~~~~~~~~~~~~~~+ (38 N -44 - 8 N^2) \ln({4 \over 3}) \biggr]
{}~.}}
\smallskip
\baselineskip 24pt plus 0.2pt minus 0.2pt

{}From \eonen, \etwopieces, \nusquare, and \rescomp\s one obtains the following
final results:

\baselineskip 12pt plus 0.2pt minus 0.2pt
\eqnla\res\resa{
E_{\Omega}^{(1)} + E_{\Omega}^{(2)} =
{N (N-1) \over 2} \nu
{}~,}
\eqnlb\resb{
E_{+}^{(1)} + E_{+}^{(2)} =
{N (N-1) \over 2} \nu
{}~,}
\eqnlcend\resc{
E_{-}^{(1)} + E_{-}^{(2)} =
{N (N-3) \over 2} \nu
+ \ln({4 \over 3})
{3 N (N-2) \over 2}
\nu^2
{}~.}
\smallskip
\baselineskip 24pt plus 0.2pt minus 0.2pt

\noindent
$E_{\Omega}$ and $E_{+}$ are among the exactly known eigenenergies of
the $N$-anyon in harmonic potential problem (see Ref.[\cho]),
and they are in perfect agreement with \resa\s and \resb.
$E_{-}$ is not known exactly, but,
for the special cases $N=3$ and $N=4$, $E_{-}^{(1)} + E_{-}^{(2)}$ has been
calculated in Refs.[\mit,\papsix], using the regularization method I,
and in Refs.[\verba,\murthy], using numerical methods;
Eq.\resc, for the
corresponding values of $N$, is in agreement with those results.

In conclusion,
the $\delta$-function regularized bosonic end perturbation theory
appears to be a reliable tool for the investigation of anyon quantum
mechanics.
I have shown explicitly for various anyon
states that its results are in agreement with the exact analytic
eigensolutions,  when such eigensolutions are known, and with the
numerical analysis of
the few anyon spectra presented in Refs.[\verba,\murthy].

This bosonic end perturbative approach may be preferable to the ones
previously introduced in the literature
because of the more physically intuitive regularization procedure.
Moreover, in some
instances properties of the matrix elements of the
$\delta$-function regularized Hamiltonian
can be used to simplify the calculations.

Indeed, by exploiting
properties of some matrix elements of the
$\delta$-function regularized Hamiltonian for $N$ anyons in an
harmonic potential, I obtained the first direct
evidence (Eq.\resc) of
the fact that there are many-anyon eigenenergies with nonlinear dependence
on the statistical parameter $\nu$. [N.B. All the presently
exactly known anyon eigenenergies$^{[\wil,\tru,\cho]}$ depend linearly
on $\nu$,
and the existence of eigenenergies with nonlinear dependence
on $\nu$ had previously been shown
only for some 3 and 4 anyon systems$^{[\cho,\mit-\murthy]}$.]

The presence of a factor $ln({4 \over 3})$ in Eq.\resc\s supports
the conjecture made in Refs.[\mit,\papsix]
that such factors should characterize a part of the general $N$-anyon
spectra.
A further investigation of the role and the physical meaning,
if there is any,
of this factor $ln({4 \over 3})$ would be very
interesting; in particular, as it has already been
suggested$^{[\ref\cap{A. Cappelli, C.A. Trugenberger, and
G.R. Zemba, Nucl. Phys. B396 (1993) 465.}]}$
for other ``noticible factors''
which have appeared in the study of anyons,
they might be related to presently unidentified
symmetries of the problem.
Eq.\resc\s is also consistent with the ansatz
proposed in Refs.[\mit,\papsix]  for the unknown portion of the spectra,
and this should encourage attempts to use that ansatz in the search
for new exact eigensolutions.

$~$

I thank S.-Y. Pi for many useful discussions on this subject,
R. Jackiw for suggesting that the results of Refs.[\mit,\papsix]
might be related to the ones of Ref.[\lozanob],
and O. Bergman and G. Lozano for conversations concerning Ref.[\lozanob].
I also thank J. Negele for hospitality.
I acknowledge support from the Istituto Nazionale di Fisica Nucleare,
Frascati, Italy.
Part of the research reported in this paper was done
at the Department of Physics, Boston University.

\vfill
\eject
\newsec{Note Added}
After this letter had been submitted for pubblication,
the Center for Theoretical Physics received a preprint
of Manuel and Tarrach$^{[\ref\tarr{C. Manuel and R. Tarrach,
Universitat de Barcelona preprint UB-ECM-PF-19/93, September 1993.}]}$
in which, in a different calculation, an analogous
$\delta$-function regularization is proposed. The combination
of my results with the ones of Ref.[\tarr] suggests that this
regularization procedure might be useful in the study of several problems.

\vfill
\eject
\vfill

\vfill\eject\immediate\closeout\rfile
\centerline{{\bf References}}\bigskip{
\catcode`\@=11\escapechar=` %
\input refs.tmp\vfill\eject}
\end